%% file: main.tex
\documentclass[a4paper,11pt]{article}

\usepackage{pos}
\usepackage{siunitx}
\usepackage[capitalise]{cleveref}
\usepackage{lineno}

\newcommand{\simu}{\texttt{T\footnotesize{AMBO}\normalsize{Sim}}}

\title{Simulation and Performance Studies for the Tau Air-Shower Mountain-Based Observatory}
\ShortTitle{TAMBO Simulation}

\author[a]{Carlos A. Arg\"{u}elles}
\author*[b]{Jeffrey Lazar}
\author[a]{William Thompson}
\author[a]{Pavel Zhelnin}

\affiliation[a]{Department of Physics and Laboratory for Particle Physics and Cosmology, Harvard University, Cambridge, MA 02138, USA}

\affiliation[b]{Université Catholique de Louvain, Pl. de l'Université 1, 1348 Ottignies-Louvain-la-Neuve}

\emailAdd{carguelles@g.harvard.edu}
\emailAdd{jeffrey.lazar@uclouvain.be}
\emailAdd{pzhelnin@g.harvard.edu}
\emailAdd{will\_thompson@g.harvard.edu}

\abstract{
While IceCube’s detection of astrophysical neutrinos at energies up to a few PeV has opened a new window to our Universe, much remains to be discovered regarding these neutrinos’ origin and nature.
In particular, the difficulty of differentiating electron- and tau-neutrino charged-current (CC) events limits our ability to measure precisely the flavor ratio of this flux.
The Tau Air-Shower Mountain-Based Observatory (TAMBO) is a next-generation neutrino observatory capable of producing a high-purity sample of tau-neutrino CC events in the energy range from \SI{1}{PeV}--\SI{100}{PeV}, i.e. just above the IceCube measurements.
An array of water Cherenkov tanks and plastic scintillators deployed in the Colca Canyon will observe the air-shower produced when a tau lepton, produced in a tau-neutrino CC interaction, emerges from the opposite face and decays in the air.
In this contribution, I will present the performance studies for TAMBO---including the expected rates, effective areas, and discrimination potential---as well as the simulation on which these studies are based.
}

\ConferenceLogo{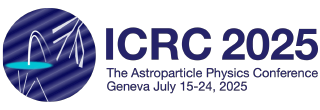}

\FullConference{39th International Cosmic Ray Conference (ICRC2025)\\
 15–24 July 2025\\
Geneva, Switzerland\\}

\begin{document}
\maketitle

\input{introduction}

\input{physics_overview}

\input{output}

\input{performance_studies}

\setlength{\bibsep}{0pt plus 0.3ex}
\bibliographystyle{unsrt}  
\bibliography{main}

\clearpage
\input{tambo-authorlist-icrc2025}

\end{document}

%% file: introduction.tex
\section{Introduction}
\label{sec:introduction}

Over the past twenty years, neutrino astronomy has evolved from the initial detection of neutrinos into a fully established field of scientific discovery.
Most of the currently active neutrino observatories belong to the class of water- or ice-Cherenkov detectors, which identify neutrinos through the Cherenkov radiation emitted by secondary particles produced in neutrino interactions.
While these instruments have played a foundational role in the development of the field, they also face several limitations.
Chief among these are challenges in distinguishing between neutrino flavors and reduced sensitivity to diffuse neutrino fluxes at energies above $\approx\SI{10}{PeV}$, due to the immense detection volumes required.

To overcome these limitations, next-generation neutrino telescopes are adopting novel detection paradigms. These approaches aim to observe either the particle cascades or radio emissions produced by neutrinos with energies $\gtrsim\SI{10}{PeV}$.
Advancing these new strategies calls for the development of specialized software tools to evaluate the scientific capabilities of upcoming experiments. For instance, the GRAND collaboration, which plans to deploy a 20,000-antenna array, has recently released a public simulation package.
However, a gap remains: no dedicated simulation framework currently exists for experiments targeting particles generated by Earth-skimming neutrinos.

The Tau Air-Shower Mountain-Based Observatory (TAMBO) is a proposed next-generation neutrino detector designed to observe tau neutrinos in the \SI{1}{PeV}--\SI{100}{PeV} energy range.
TAMBO is specifically optimized to achieve high-purity measurements of astrophysical tau neutrinos by leveraging the relative scarcity of atmospheric tau neutrinos at these energies.
Unlike electron- or muon-neutrinos, tau-neutrinos at PeV energies can produce extensive air showers (EASs) through the decay of the secondary tau lepton.
These upward- or horizontally-propagating showers serve as distinctive signatures of tau-neutrino interactions.
To distinguish them from cosmic-ray-induced showers, TAMBO exploits the fact that tau-neutrinos can produce signals only after traversing several kilometers of rock, which effectively suppresses the cosmic ray background.
The observatory will maximize sky coverage by placing an array of water-Cherenkov and plastic scintillator detectors along the slope of a deep canyon.

This proceeding introduces \simu{}, a \texttt{Julia}-based software package developed to simulate the physics underpinning TAMBO.
Although TAMBO shares certain design elements with both traditional neutrino detectors and ground-based air-shower arrays, its unique geometry and detection method necessitate a new simulation framework.
Earth-skimming neutrino trajectories and the inclined placement of the detector necessitate the accurate modeling of local terrain, unlike in many neutrino simulations that assume a simplified spherical Earth. 
Furthermore, unlike conventional air-shower arrays designed to detect vertical showers on flat terrain, TAMBO targets upward-going showers with a detector deployed on steep canyon walls.
These key differences underscore the need for a dedicated simulation approach, which TAMBOSim is designed to provide.

%% file: physics_overview.tex
\section{Overview of Simulated Physics}
\label{sec:physics_overview}

\begin{figure}
    \centering
    \includegraphics[width=1.0\linewidth]{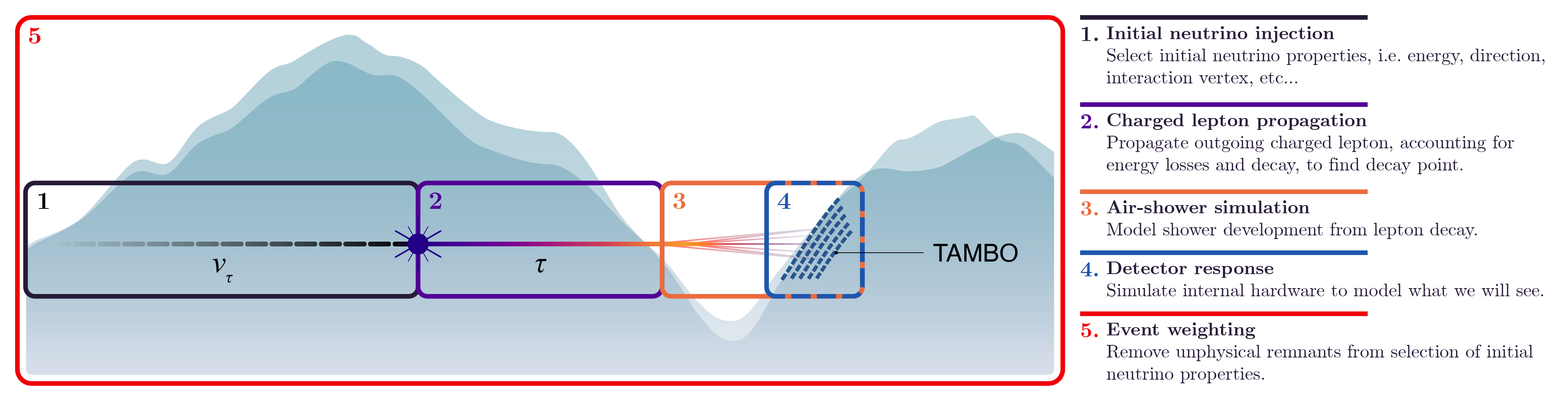}
    \caption{\textbf{\textit{Simulated physical processes.}}
    Each box represents a discrete step in the simulation chain.
    }
    \label{fig:physics-overview}
\end{figure}

\simu{}'s core functionalities can be separated into five stages: initial event injection, charged lepton propagation and decay, air-shower simulation, detector response, and event weighting.
These stages are shown schematically in \cref{fig:physics-overview}.
This section reviews the relevant physics for each step and provides a qualitative description of our approach.
Furthermore, we will also attempt to highlight other approaches in neutrino astronomy when applicable.

In the initial particle injection, the tau neutrino and charged tau lepton properties---energy, direction, and interaction vertex---are chosen.
This step should balance two countervailing needs: covering all phase space that could lead to detection and maintaining computational efficiency.
In the case of the first two quantities, we sample the tau neutrino's energy at Earth's surface from a power law and the direction uniformly on the surface of a unit sphere.
If necessary, the neutrino is propagated to the simulation region, including the effects of tau regeneration, to get the neutrino's energy near the detector.
There is no shortage of approaches for selecting the neutrino interaction vertex; see~\cite{osti_5884484,Hill:1996hzh,Gazizov:2004va,Yoshida:2003js,Bailey:2002,IceCube:2020tcq} for various examples.
Our approach most closely resembles the so-called ranged injection developed in \cite{IceCube:2020tcq}, which dynamically accounts for the energy-dependent distance that charged leptons can travel between the interaction vertex and the detector.

Once these initial properties have been set, the charged tau lepton can be propagated through the environment to the point of decay.
In this work, \simu{} relies on the \texttt{PROPOSAL} library to propagate and decay the charged tau leptons.
This includes up-to-date cross sections for ionization, bremsstrahlung, photonuclear interactions, electron pair production, the Landau–Pomeranchuk–Migdal effect, and the Ter-Mikaelian effect.
It is worth noting that while charged tau leptons produced in high-energy weak interactions are nearly 100\% polarized, \texttt{PROPOSAL} assumes unpolarized charged tau leptons.
This can indeed affect the energy of the decay products~\cite{Arguelles:2022bma}; however, we have taken some steps to mitigate this effect.

Suppose the charged tau lepton decays hadronically or electromagnetically in air. In that case, these decay products can generate an EAS, whose detailed particle content must be propagated to the face of the mountain.
This includes accounting for all particle creation and decay, interactions with the surrounding air, and deflection due to the Earth's magnetic field.
\simu{}'s use the \texttt{CORSIKA} package to simulate EASs.
In particular, we use \texttt{CORSIKA8}~\cite{CORSIKA:2023jyz} to simulate up-going EASs projected on the inclined plane that represents the mountain in this simulation.

Once the particles from an EAS have reached the mountain, we must determine which, if any, of them passed through a detection module.
We determine whether the module registers each particle by computing the photoelectron yield, which varies across particle species and as a function of energy.
Finally, we determine whether a particular event triggered the detector by examining the distribution of triggered modules across the entire detector.

These MC events are then converted to a physical rate by calculating weights.
This requires computing the probability of being injected with a particular parameter set and the probability of emerging from the desired physical distribution.
The ratio of the latter to the former gives a quantity we will call the event's \texttt{oneweight}.
This quantity is proportional to the detector acceptance and can easily be converted to a rate by multiplying by a flux of neutrinos.

%% file: output.tex
\section{Output Data Structures}
\label{sec:output}

\begin{figure}[t]
    \centering
    \includegraphics[width=0.95\linewidth]{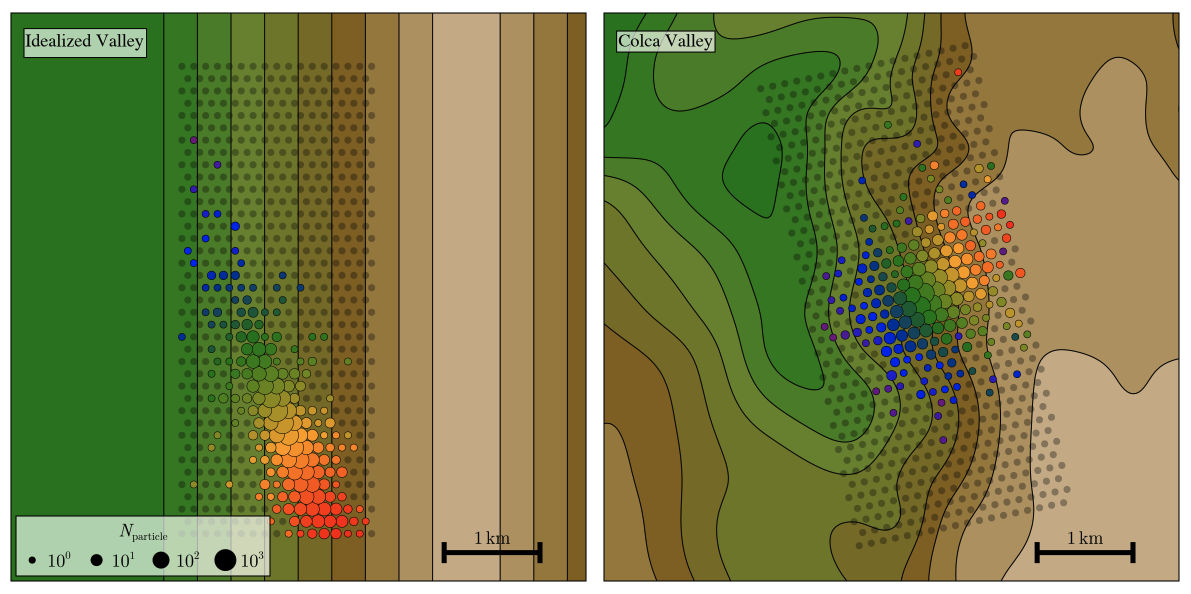}
    \caption{
    \textbf{\textit{Event displays in two geometrical configurations.}}
    The left panel shows a straight-walled, ``idealized valley'' similar to that used in a previous study.
    The right panel shows an event in the Colca Valley, centered on $15.639^{\circ}$ S, $72.165^{\circ}$ W.
    Each dot represents a detection module.
    Triggered modules are colored, whereas untriggered modules are left gray.
    The color scale gives the time at which the first particle arrived at the detection module---red is early and purple is late---and the size of the circle indicates the number of particles detected at a particular module.
    }
    \label{fig:events}
\end{figure}

The data created by \simu{} must be stored in a data format that satisfies several conditions.
First and foremost, the format must be interoperable, i.e., it must be readable across various operating systems and programming languages.
Second, the format must strike a balance between disk usage, efficient I/O, and user ease of use.
In light of these constraints, \simu{} takes a hybrid approach and serializes output data as either \texttt{Arrow} or \texttt{Parquet} files.
Both formats are columnar data structures; however, each is optimized for different tasks. \texttt{Parquet} files are optimized for efficient disk storage at the cost of computation overhead when encoding and decoding.
On the other hand, \texttt{Arrow} files are optimized for in-memory computation and allow for the serialization of custom types.
This means that structures defined in \simu{} can be saved directly and recreated if the file is opened in Julia.
If opened in a language other than \texttt{Julia}, the data will appear as a nested set of arrays.

The simulation is staged according to the steps outlined in \cref{fig:physics-overview}, with the option to write the output at each stage.
The data generated by the initial neutrino injection, charged lepton propagation, and detector response is relatively modest relative to that generated by the air-shower simulation.
In the latter step, we utilize the \texttt{Parquet} format to leverage the additional data compression.
In the former cases, we employ \texttt{Arrow} to benefit from the additional structure-preserving functionality.
Furthermore, since the event weighting is fast and application-specific, this data is not stored in an output file.
In the following paragraphs, we describe the format of the output that is generated at each step.
All output information is reported in natural units, i.e., in a unit system where $c=\hbar=1$.

The \texttt{Arrow} file produced by the initial neutrino injection step contains information about the initial tau neutrino at the surface of the Earth and upon entering the \simu{} region.
In general, the particle state can differ due to energy losses and scattering endured during propagation through the Earth.
In addition to information about the initial neutrino, this file also contains information about the initial state of the charged tau lepton created in the CC interaction.
The state of each particle follows the \texttt{Particle} structure from \simu{}.
This structure contains information about the particle's energy, position, and direction.
Finally, this file contains a unique event identification number and the column depth at which the event was generated.
Additionally, since the event weighting only depends on the injection parameters, we include the oneweight in this \texttt{Arrow} file as well.

The \texttt{Arrow} file produced after propagating the final-state tau lepton contains the state of the lepton after propagation.
Additionally, this file contains any daughter particles produced if the charged tau lepton decays.
In both cases, this information is encoded in \texttt{Particle} structs.
The file also contains a boolean specifying if the tau lepton decayed.
This value will be false if the tau lepton reaches the edge of the \simu{} simulation region before decaying.
Finally, the file contains information about the energy losses the charged tau lepton underwent.
These are broken down into stochastic and continuous losses.
The former is a list of interactions in which the energy loss is greater than a specified value, while the latter is the sum of all energy losses lower than the specified value.
Since the initial neutrino injection and charged tau lepton propagation are fast relative to the next step, these are typically run together.
Thus, the information mentioned above may appear in the same \texttt{Arrow} file.

As discussed previously, the large number of particles---sometimes on the order of millions---created in the air-shower simulation motivates storing the output in \texttt{Parquet} files.
These files follow the default format of the \texttt{CORSIKA} package, and store the particle type, energy, position, and momentum of all particles that reach the readout plane or planes.
In addition to the main \texttt{Parquet} file, \texttt{CORSIKA} also generates several ancillary files that may be useful.
The \texttt{asdlfjasdjkfhadskjfh.txt} specifies the \texttt{CORSIKA} configuration, including the transformation between \texttt{CORSIKA} and \simu{} coordinates.
Typically, this ensemble of files is created for each decay product that contributes to the air shower.
This is done because running \texttt{CORSIKA} is highly computationally expensive, requiring a high degree of parallelization.

After simulating the detector response, another \texttt{Arrow} file is created containing the data.
This data takes the form of a hash map mapping detection module identification numbers to lists of particles that reached that detection module.
These maps contain all relevant data---time of arrival, number of particles---to create event displays.
See \cref{fig:events} for examples of event displays in an idealized valley as well as an example location in the Colca Valley.
These are typically stored before triggering because constructing these maps requires looping over all particles in an air shower to determine if they intersect any module.
While this is not as computationally intensive as the air-shower simulation itself, it is significantly more expensive than the triggering step itself.
The primary cause of this is the large air showers with many particles, as well as the overhead associated with reading the requisite \texttt{Parquet} file into memory and looping over all particles.
Thus, we can investigate multiple triggering schemes with the same map by writing at this stage.

%% file: performance_studies.tex
\section{Performance Studies}
\label{sec:performance}

\begin{figure}[t]
    \centering
    \includegraphics[width=0.6\linewidth]{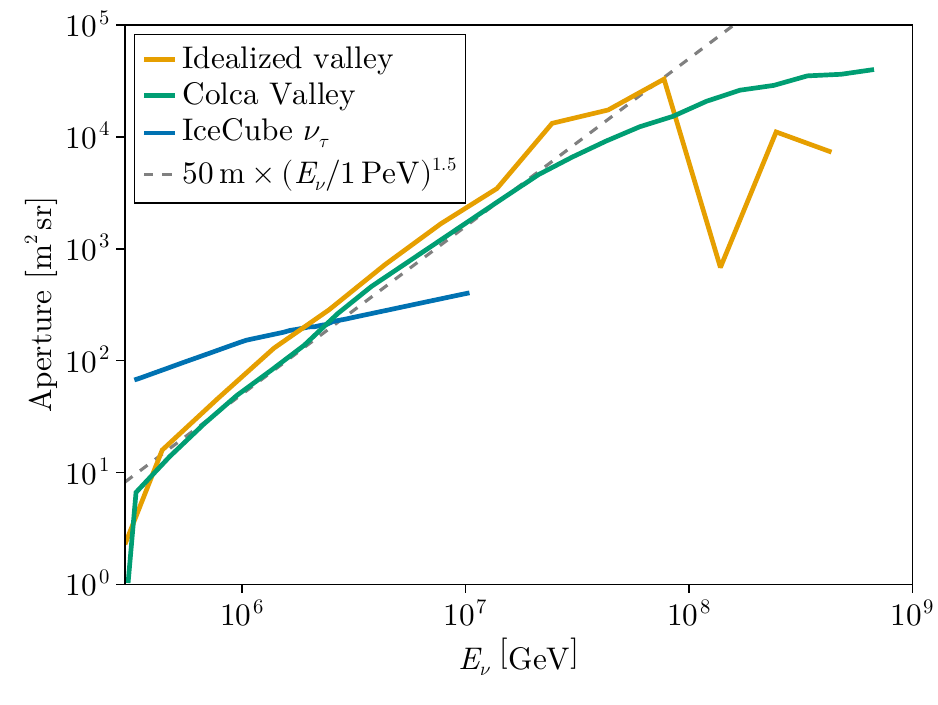}
    \caption{
    \textbf{\textit{Aperture for a 5,000-module detector deployed in both the Colca Valley and in the previously described idealized valley.}}
    The idealized valley achieves a slightly higher exposure across most energies due to the larger average distance between its sides, which allows more time for the EAS to develop.
    These are compared to the IceCube's tau-neutrino aperture, which becomes subdominant at $\approx\SI{3}{PeV}$.
    Additionally, a line showing the expected $E_{\nu}^{1.5}$ scaling is plotted behind these lines.
    The apertures are in good agreement with this scaling until the highest energies, where the tau lifetime exceeds the width of the valley.
    The fluctuations in the idealized valley curve are due to limited Monte Carlo statistics in this configuration.
    }
    \label{fig:apertures}
\end{figure}

Since TAMBO does not have $2\pi$ azimuthal field of view, one cannot readily compare effective areas directly.
Thus, the figure by which we evaluate the performance is the detector aperture, i.e., the effective area multiplied by the portion of the sky that the detector sees at a given time.
This allows for direct comparison between different types of observatories.

In \cref{fig:apertures}, we compare the apertures obtained from \simu{} for 5,000-module detectors deployed both at the nominal location within the Colca Valley and in the straight-walled idealized valley shown in \cref{fig:events}.
The aperture obtained by the detector in the idealized valley is generally higher than that for the Colca-Valley-based detector.
This can be understood because the average distance between the idealized valley's walls is larger than that in the Colca Valley.
Thus, EASs in the idealized valley arrive at the opposite side of the valley closer to the shower maximum, and the increased particle yield improves the chance of triggering on a shower.
These should be compared to the dashed line that shows $E_{\nu}^{1.5}$ scaling.
This scaling is expected as the neutrino cross section scales like $E_{\nu}^{0.5}$ at these energies and the tau lepton range scales like $E_{\nu}$, enabling interactions in more rock.

In addition to the comparison between the two valley geometries, we also compare to IceCube's tau-neutrino aperture.
In both geometries, TAMBO's aperture overtakes IceCube's at neutrino energies of $\approx\SI{3}{PeV}$.
This is the regime where IceCube has very few events, highlighting the complementarity between Cherenkov neutrino telescopes and TAMBO.
Furthermore, the IceCube aperture stops at \SI{10}{PeV} due to the negligible event rate in a detector the size of IceCube.

%% file: tambo-authorlist-icrc2025.tex
\section*{Full Author List: TAMBO Collaboration}

\scriptsize
\noindent
Carlos A. Argüelles$^{1}$,
José Bazo$^{2}$,
Christopher Briceño$^{3}$,
Mauricio Bustamante$^{4}$,
Saneli Carbajal$^{5}$,
Víctor Centa$^{6}$,
Jaco de Swart$^{7}$,
Diyaselis Delgado$^{1}$,
Tommaso Dorigo$^{8,9}$,
Anatoli Fedynitch$^{10}$,
Pablo Fernández$^{11}$,
Alberto Gago$^{2}$,
Alfonso García$^{12}$,
Alessandro Giuffra$^{3}$,
Zigfried Hampel-Arias$^{13}$,
Ali Kheirandish$^{14,15}$,
Jeffrey P. Lazar$^{16}$,
Peter M. Lewis$^{17}$,
Daniel Menéndez$^{6}$,
Marco Milla$^{6}$,
Alberto Peláez$^{3}$,
Andres Romero-Wolf$^{18}$,
Ibrahim Safa$^{19}$,
Luciano Stucchi$^{5}$,
Jimmy Tarrillo$^{3}$,
William G. Thompson$^{1}$,
Pietro Vischia$^{20}$,
Aaron C. Vincent$^{21,22,23,1}$,
Pavel Zhelnin$^{1}$

\vspace{1em}
\noindent
$^{1}$ Department of Physics \& Laboratory for Particle Physics and Cosmology, Harvard University, Cambridge, MA, USA \\
$^{2}$ Sección Física, Departamento de Ciencias, Pontificia Universidad Católica del Perú, Lima, Perú \\
$^{3}$ Electrical and Mechatronic Department, Universidad de Ingenieria y Tecnologia, Barranco, Perú \\
$^{4}$ Niels Bohr International Academy, Niels Bohr Institute, University of Copenhagen, Denmark \\
$^{5}$ Universidad del Pacífico, Lima, Perú \\
$^{6}$ Instituto de Radioastronomía, Pontificia Universidad Católica del Perú, Lima, Perú \\
$^{7}$ Department of Physics \& Program in Science, Technology, and Society, MIT, Cambridge, MA, USA \\
$^{8}$ Luleå University of Technology, Luleå, Sweden \\
$^{9}$ INFN, Sezione di Padova, Italy \\
$^{10}$ Institute of Physics, Academia Sinica, Taipei City, Taiwan \\
$^{11}$ Donostia International Physics Center DIPC, San Sebastián, Spain \\
$^{12}$ Instituto de Física Corpuscular (IFIC), CSIC and Universitat de València, Spain \\
$^{13}$ Los Alamos National Laboratory, Los Alamos, NM, USA \\
$^{14}$ Department of Physics \& Astronomy, University of Nevada, Las Vegas, NV, USA \\
$^{15}$ Nevada Center for Astrophysics, University of Nevada, Las Vegas, NV, USA \\
$^{16}$ CP3, Université catholique de Louvain, Louvain-la-Neuve, Belgium \\
$^{17}$ Department of Physics and Astronomy, University of Hawaii at Manoa, Honolulu, HI, USA \\
$^{18}$ Jet Propulsion Laboratory, California Institute of Technology, Pasadena, CA, USA \\
$^{19}$ Department of Physics, Columbia University, New York, NY, USA \\
$^{20}$ Calle San Francisco 3, Oviedo, Principado de Asturias, España \\
$^{21}$ Department of Physics, Engineering Physics and Astronomy, Queen's University, Kingston, ON, Canada \\
$^{22}$ Arthur B. McDonald Canadian Astroparticle Physics Research Institute, Kingston, ON, Canada \\
$^{23}$ Perimeter Institute for Theoretical Physics, Waterloo, ON, Canada

\subsection*{Acknowledgments}

\noindent
We would also like to thank the Milton Family Fund at Harvard, the Harvard Faculty of Arts and Sciences Dean's Fund for Promising Scholarship, and the Harvard-UTEC fund.  
The initial development of the social aspects of a large neutrino telescope in the Peruvian Andes was supported by the Radcliffe Institute for Advanced Study at Harvard University.  
CAD and WT were partially supported by the Canadian Institute for Advanced Research (CIFAR) Azrieli Global Scholars program through this work.  
CAA is supported by the Faculty of Arts and Sciences of Harvard University, the National Science Foundation (NSF), the NSF AI Institute for Artificial Intelligence and Fundamental Interactions, the Research Corporation for Science Advancement, and the David \& Lucile Packard Foundation.  
CAA and JL were supported by the Alfred P. Sloan Foundation for part of this work.  
PV is supported by the ``Ramón y Cajal” program under Project No. RYC2021-033305-I funded by the MCIN/AEI/10.13039/501100011033 and by the European Union NextGenerationEU/PRTR.  
MB is supported by \textsc{Villum Fonden} under project no.~29388.  
AG is supported by the CDEIGENT grant No. CIDEIG/2023/20.  
JL is a postdoctoral researcher at the Fonds de la Recherche Scientifique -- FNRS.  
JB and AG acknowledge the Dirección de Fomento de la Investigación (DFI-PUCP) for funding under grant CAP-PI1144.  
JdS is supported by the American Institute of Physics Robert H.G. Helleman Memorial Postdoctoral Fellowship.